\RequirePackage[2020-02-02]{latexrelease}
\documentclass[aps,preprint,12pt]{revtex4}%
\usepackage{amsfonts}
\usepackage{amsmath}
\usepackage{amssymb}
\usepackage{graphicx}%
\providecommand{\U}[1]{\protect\rule{.1in}{.1in}}
\begin{document}
	\preprint{ }
	\title[ ]{Information versus Physicality: On the Nature of the Wavefunctions of Quantum Mechanics}
	\author{C. S. Unnikrishnan}
	\affiliation{School of Quantum Technology, \\The Defence Institute of Advanced Technology,
		Pune 411025, India}
	
	\begin{abstract}
		
The physical states of matter and fields are represented in the quantum theory with complex valued wavefunctions, or more generally by  quantum states in an abstract linear vector space. Determining the physical nature of wavefunctions remains an open problem that is at the very core of quantum mechanics,  About a decade ago, Pusey, Barrett and Rudolf (PBR) claimed to prove an ontologically real status of wavefunctions by ruling out  $\psi$-epistemic models. The result was obtained by associating wavefunctions to hypothetical distributions of notional physical states, and by examining whether some physical states were associated with more than one wavefunction, a criterion they chose for defining a wavefunction as `epistemic'. I show that the starting assumption in the PBR argument, of associating a wavefunction with a distribution of physical states, is flawed and contradictory to the linear structure of quantum mechanics coupled with its quadratic Born's rule.  Since none of the axioms or calculations of observable statistical results in the standard quantum theory depends on specifying the physical nature of a $\psi$-function, the considerations in the PBR paper, involving a standard process of the preparation and projective measurements of quantum states,  cannot address the ontological status of the wavefunctions in space and time. 		
	\end{abstract}
	\startpage{1}
	\endpage{102}
	\maketitle
	%\tableofcontents

\section{Introduction}
Quantum mechanics that was formulated nearly a century ago is a mature theory with unprecedented reach and scope. Its formalism based on the Schr\"odinger equation represents the notional physical state of a system (or particle) by a complex-valued wavefunction, also called a ``$\psi$-function''. Wavefunctions are related to  a more abstract notion of a quantum state, represented as a normalized vector in a linear vector space. While the absolute squared value of the wavefunction is identified as a probability density (Born's rule), there is no consistent  physical understanding of the wavefunction itself. This serious lacuna is at the core of several unsolved issues of quantum theory, like the problem of the collapse of the quantum state during an observation, and the quantum measurement problem. 

The question whether a  wavefunction that represents a quantum physical state has an ontological existence in space and time (labelled an ``ontic'' state by some authors), or whether it represents merely the state of an observer's knowledge about a physical state, and hence only of an ``epistemic'' status, has been debated since Schr\"odinger's influential review paper in 1935~\cite{Sch-cat}. Schr\"odinger discussed in detail the difficulty of ascribing an ontological status to the wavefunction, forcing the ``rejection of realism'', and he considered an interpretation of the wavefunction as ``a catalogue of expectations''. In the initial era of quantum mechanics, there was a belief that a wavefunction had a physical existence, expressed as the wave-particle duality of matter. However, this belief waned quickly, because such an interpretation could not be maintained consistently, especially for multi-particle wavefunctions~\cite{Sch-cat,deBroglie,JvN-book}. (The naive and inaccurate identification of the wavefunction with a real `matter-wave' in space is still widespread, though.) 

The question whether a $\psi$-function is physically real or not is factually irrelevant for any calculation or quantitative prediction of the quantum theory, because the axioms of the theory as well as the well defined mathematical formalism to obtain the observable statistical results are entirely independent of this consideration~\cite{JvN-book,Dirac}. However, a physical understanding of the theory and a causal explanation of quantum phenomena  are admittedly related to unravelling the nature of $\psi$-functions. 

The different views on the nature of $\psi$-functions are straddled by two extreme views: one is that a wavefunction has the physical reality of an ontological existence in space and time (ontic), and the other is that a wavefunction is merely a calculational tool and an information catalogue about the quantum system that represents the knowledge (epistemic), possibly incomplete, of an observer~\cite{Fuchs}. Then there are middle grounds that combine aspects from these two extreme views~\cite{Brown,Hance,Hubert,Sabine}.

In a paper published in 2012~\cite{PBR},  M. F. Pusey, J. Barrett and T. Rudolf (PBR)  claimed to have proved a no-go theorem: ``if the quantum state merely represents information about the real physical state of a system, then experimental predictions are obtained which contradict those of quantum theory. The argument depends on a few assumptions. One is that a system has a `real physical state', not necessarily completely described by quantum theory, but objective and independent of the observer. This assumption only needs to hold for systems that are isolated, and not entangled with other systems... The other main assumption is that systems that are prepared independently have independent physical states.'' Though PBR admitted that the main assumption would be denied outright by  those who uphold a solely epistemic status of $\psi$-functions, as a calculational tool, they stated their result with an implied universal applicability, as ``$\psi$-epistemic models cannot reproduce the predictions of quantum theory''. However, the PBR result was actually about a very restricted   $\psi$-epistemic model, defined as one in which the distributions of hypothetical physical states overlap at least for a pair of $\psi$-functions, making it impossible to uniquely point to a $\psi$- function, given a physical state. If no pair of distributions overlaps, then the $\psi$-functions are defined as ontic and physically real. 

It is important to note that any model that considers the wavefunction as an information catalogue and a mathematical device to predict all statistical results of the quantum theory, without any further attempt to associate it with hypothetical physical states etc., are totally out of the purview of the PBR argument. Such a model is clearly a $\psi$-epistemic model, without added speculations or  attempts to describe an unobservable physicality `behind' quantum mechanics. With only Born's rule, the model will agree with all statistical predictions of quantum mechanics. Therefore, it is obvious that the PRB general claim that $\psi$-epistemic models contradict the predictions of quantum theory has no general validity.

The PBR claim created a persistent impression that one could prove the definite ontological reality of $\psi$-functions by first postulating a model outside quantum mechanics, with an unverifiable association between a $\psi$-function and the notional physical states it supposedly represents, and then by examining whether some aspects of such a hypothetical association contradict the verifiable  predictions of quantum mechanics. 

In the standard quantum theory, two $\psi$-functions that differ by more than a pure phase factor represent two distinct physical states, irrespective of any interpretation of the $\psi$-function. In particular, quantum mechanics does not allow a physical state to be represented by two distinct wavefunctions, which should be distinguished from the fact that  a $\psi$-function expressed in one eigen-basis can give different results stochastically when measured in another basis. The Hermitian operators for observables and projective measurements are mathematically and physically well-defined notions that are independent of whether a $\psi$-function is ontological or epistemic. From the preparation of a quantum state and its evolution in space and time to a projective measurement in any orthogonal basis, only the quantum state ($\psi$-function) and the operators representing physical variables figure in the calculations of the observable statistical results; no reference to any underlying physical state is ever needed. In other words, the nature and interpretation of the wavefunction, beyond what is stated in  Born's rule, is entirely irrelevant for any ensemble averaged statistical results of quantum mechanics~\cite{JvN-book,Dirac}. Besides, there is the method of matrix quantum mechanics that does not directly refer to a wavefunction or the wave-particle duality, but which is formally equivalent in reproducing the observable statistical results. \emph{A set of physical operations consisting of the preparation of a quantum state and its projective measurements will give statistical results that are insensitive to whether the $\psi$-function is ontic or epistemic}. In other words, there is no factor that distinguishes between an ontic and epistemic $\psi$-function in the entire structure of the standard quantum theory. Moreover, the distinction of $\psi$-functions as `physically real' and `epistemic' in the PBR paper was based merely on their  restricted definitions of two mutually exclusive properties of the postulated association in a specific model~\cite{PBR,Harrigan,JLuc}. There have been many works that analysed the PBR model and argument, and there have been some attempted experiments along the PBR reasoning~\cite{Hardy-Pus,Fine-Pus,Nigg-exp,Ringbauer-exp,Leifer,Carcassi,Calosi,Wallegham}. It is obvious that any inconsistency in their central postulate, of an association between a wavefunction and a hypothetical distribution of physical states, will render the PBR argument invalid. We will see, with multiple proofs, that this is indeed the case.

\section{Methods and Results}
\subsection{The Fundamental Fault in the PBR Model}\label{sec2}
A central postulate in the PBR paper is that a distribution labelled by the quantum state $\psi$, of hypothetical real physical states $\lambda$, is associated with a quantum state ($\psi$-function)~\cite{PBR}. Thus, a wavefunction $\psi$ is mapped to a distribution of physical states $\mu_\psi(\lambda)$. 

In quantum mechanics, the set of quantum states $e^{i\varphi}|\psi\rangle$ belong to an equivalence class, representing the same quantum physical state. All these states give identical predictions for the statistical results of projective measurements. A measuring instrument cannot distinguish between $|\psi\rangle$, $-|\psi\rangle$, and  $i|\psi\rangle$, for example.

The physical nature of the relation between a quantum state $|\psi\rangle$ and the associated physical states is explicitly specified by PBR, and clearly illustrated in  the figure 2 of their paper. \emph{The output of a device for the preparation of a quantum state $|\psi\rangle$ is indeed a  physical state from the associated distribution} (figure 1a). Such a physical state can be subjected to a (projective) measurement with another device arranged in a suitable basis ${|\phi_i\rangle}$. According to PBR, `the outcome of a measurement can only depend on the physical state ... at the time of measurement.' We observe only the results of the measurements, and not the quantum state or the physical states. 

\begin{figure}
	\centering
	\includegraphics[width=0.8\linewidth]{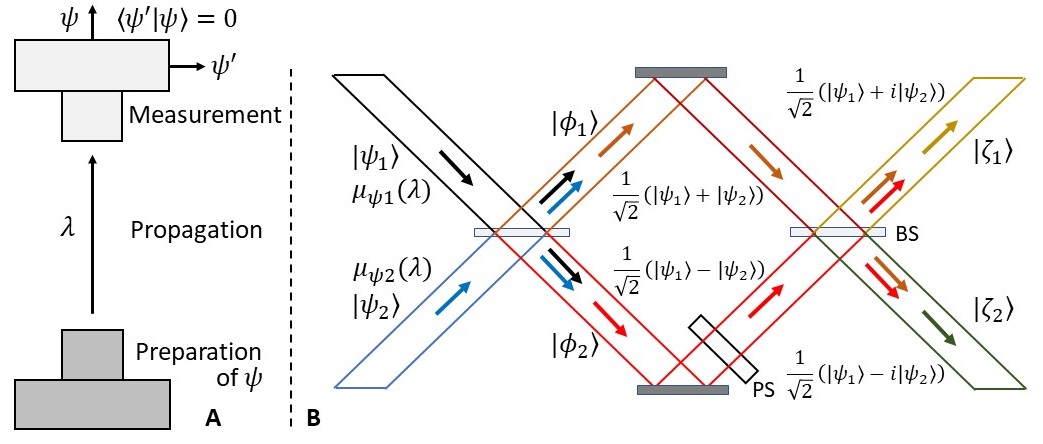}
	\caption{A) The PBR model of the stages of preparation of a quantum state $\psi$, the propagation of the associated physical state $\lambda_\psi$, and its projective measurement (as per fig. 2 in the PBR paper). B) The linear relations between different $\psi$-functions according to the quantum theory. Two orthogonal quantum states $\psi_1$ and $\psi_2$ superpose in two distinct ways at a `beam splitter' (BS) to give the orthogonal states $\phi_1=(\psi_1+\psi_2)/\sqrt{2}$ and $\phi_2=(\psi_1-\psi_2)/\sqrt{2}$. They further superpose with a relative phase (PS) to give the orthogonal states $\zeta_1=(\psi_1+i\psi_2)/\sqrt{2}$ and $\zeta_2=(\psi_1-i\psi_2)/\sqrt{2}$. But, $\psi_1$ and $\psi_2$ are originally mapped to two  non-overlapping distributions $\mu_{\psi 1}$ and $\mu_{\psi 2}$ of real physical states in the PBR model, whereas there cannot be even one physical state that is common to $\phi_1$ and $\phi_2$, or to $\zeta_1$ and $\zeta_2$. The diagram is valid in the reverse sequence as well, $\zeta\rightarrow \phi\rightarrow\psi$.}
	\label{fig:pbr-states}
\end{figure}

Consider two mutually orthogonal $\psi$-functions $|\psi_1\rangle$ and $|\psi_2\rangle$, and the associated hypothetical distributions $\mu_{\psi1}(\lambda)$ and $\mu_{\psi2}(\lambda)$. Since the quantum states are orthogonal, the distributions are non-overlapping. If we superpose these quantum states to obtain a new quantum state $|\psi_+\rangle=(|\psi_1\rangle+|\psi_2\rangle)/\sqrt{2}$, the total equivalence of the $\psi$-states on both sides of the equal sign implies that the state $|\psi_+\rangle$ is mapped to some combination of the non-overlapping distributions $\mu_{\psi1}(\lambda)$ and $\mu_{\psi2}(\lambda)$ (Fig.~\ref{fig:pbr-states}B). This is consistent with the results of projective measurements in the basis $\{|\psi_1\rangle,|\psi_1\rangle\}$. 
Another superposition of the same quantum states is $|\psi_-\rangle=(|\psi_1\rangle-|\psi_2\rangle)/\sqrt{2}$. Then the state $|\psi_-\rangle$ is mapped to some combination of the same two non-overlapping distributions $\mu_{\psi1}(\lambda)$ and $\mu_{\psi2}(\lambda)$. Of course, the state $|\psi_-\rangle$ gives the same results for projective measurements in the basis $\{|\psi_1\rangle,|\psi_2\rangle\}$, as the state $|\psi_+\rangle$.  \emph{However, there cannot be even one physical state that is common between $|\psi_+\rangle$ and $|\psi_-\rangle$ because these two states are mutually orthogonal}!

\begin{multline}
	\langle \psi_-|\psi_+\rangle=\frac{1}{\sqrt{2}}(|\psi_1\rangle+|\psi_2\rangle)\cdot \frac{1}{\sqrt{2}}(|\psi_1\rangle-|\psi_2\rangle)\\ 
	=\frac{1}{2}(\langle \psi_1|\psi_1\rangle-\langle \psi_1|\psi_2\rangle+\langle \psi_2|\psi_1\rangle-\langle \psi_2|\psi_2\rangle)=0	
\end{multline}
Since  PBR's central assumption leads to the inconsistent conclusion that several mutually orthogonal quantum states are mapped to at least some common ontic physical states, the assumption and  the PBR claim are invalid.

Note that there is no escape from this contradiction by postulating an altogether different distribution $\mu_{\psi+}(\lambda)$ associated with $|\psi_{+}\rangle$, and $\mu_{\psi-}(\lambda)$ with $|\psi_{-}\rangle$, etc., \emph{because of the closed linear vectorial structure of quantum mechanics} that gives $|\psi_{1}\rangle=(|\psi_{+}\rangle+|\psi_{-}\rangle)/\sqrt{2}$ and $|\psi_{2}\rangle=(|\psi_{+}\rangle-|\psi_{-}\rangle)/\sqrt{2}$, etc.  

As a familiar example, we can take the spinorial quantum states $|+z\rangle$ and $|-z\rangle$. These are mapped to the non-overlapping distributions of underlying physical states $\mu_{|+z\rangle}(\lambda)$ and $\mu_{|-z\rangle}(\lambda)$. 
Measurements on a state prepared as $|+z\rangle$ (or as $\exp^{i\varphi}|+z\rangle$), with a device set for the projection on to the state $|+z\rangle$, give the result $1$ with certainty, and the result $0$ with certainty with a measuring device set for the state $|-z\rangle$. If the state is prepared as $|-z\rangle$, exactly opposite results would be obtained. 

We have the freedom to spatially superpose the two quantum states $|+z\rangle$ and $|-z\rangle$, as $|\psi_s\rangle=(|+z\rangle+|-z\rangle)/\sqrt{2}$. Since $|+z\rangle$ and $|-z\rangle$ are mapped uniquely to $\mu_{|+z\rangle}(\lambda)$ and $\mu_{|-z\rangle}(\lambda)$, the linear structure of the theory implies that $|\psi_s\rangle$ (which is  the state $|+x\rangle$) is mapped to  the non-overlapping distributions $\mu_{|+z\rangle}(\lambda)$ and $\mu_{|-z\rangle}(\lambda)$ of physical states. This is of course compatible with what one sees in quantum measurements.  However, the inconsistency arises because we can have the superposition $|-x\rangle=(|+z\rangle-|-z\rangle)/\sqrt{2}$, which is mapped to the members of the same distributions $\mu_{|+z\rangle}(\lambda)$ and $\mu_{|-z\rangle}(\lambda)$. \emph{Thus, there are some common physical states $\lambda_c$ from the non-overlapping distributions  $\mu_{|+z\rangle}(\lambda)$ and $\mu_{|-z\rangle}(\lambda)$ mapped to both the quantum states} $|+x\rangle$ and $|-x\rangle$. This is inconsistent, and ruled out empirically, because $|+x\rangle$ and $|-x\rangle$ are orthogonal, $\langle +x|-x\rangle=0$.

The situation is much worse in detail, because there are other states  like $|+y\rangle$ and $|-y\rangle$ that are mutually orthogonal, and synthesized by linearly superposing the same base quantum states $|+z\rangle$ and $|-z\rangle$. An infinity of such mutually orthogonal quantum states gets associated with the elements of the same non-overlapping distributions of physical states $\mu_{|+z\rangle}(\lambda)$ and $\mu_{|-z\rangle}(\lambda)$, through the initial mapping of the states $|+z\rangle$ and $|-z\rangle$ to these distributions.

A measurement device set to project on the states $|\pm z\rangle$ cannot distinguish between $|+x\rangle$ and$|-x\rangle$; both give the results $\pm 1$ with equal probability. But, if we make the grave mistake of ignoring the dynamics (phases) and focus only on the results of projective measurements (and their probabilities), based on analogies with classical stochastic situations, we will end up with  irreconcilable inconsistencies~\cite{Unni-JvN}.

To see this strikingly,  consider the quantum state $|+x\rangle$ representing a spin-1/2 particle polarized along $+x$. The PBR scheme maps this state to a positive definite distribution of physical states $\mu_{+x}(\lambda)$. As mentioned explicitly in the PBR paper (see fig. 2 in their paper), a physical state $\lambda$ is what is realized in the state preparation device. We can pass it through  a double Stern-Gerlach arrangement that notionally separates the states $|+z\rangle$ and $|-z\rangle$ and then \emph{recombines after adding an extra phase} $\exp(i\pi$) to only the $|-z\rangle$ internal path, without any measurements within the device. According to quantum mechanics, and all experimental evidence, what comes out corresponds to the quantum state $|-x\rangle$ representing a spin-1/2 particle polarized along $-x$! However, in the PBR scheme, this requires the impossible transformation of the physical state to a member of the non-overlapping distribution $\mu_{-x}(\lambda)$ associated with the quantum state $|-x\rangle$.

The inconsistency in the PBR assumption of mapping a quantum state to a distribution of hypothetical physical states is related to the impossibility of reproducing quantum mechanical results with dispersion-free ensembles postulated in hidden variable theories~\cite{JvN-book,Unni-JvN}. That is not surprising because the PBR definition was inspired by a classical analogy with phase space distributions.

I have shown that the PBR claim of determining the physical nature of the wavefunction from the results of standard preparation and projective measurements of quantum states is invalid, because their starting assumption itself-- their model mapping of $\psi$-functions with a distribution of physical states-- contradicts the linear structure of quantum superpositions coupled with  its quadratic connection to observable results through Born's rule. Now we can discuss the PBR argument in more detail to emphasize its fault in multiple ways.

\subsection{Distributions Associated With Continuous Quantum States}
\begin{figure}
	\centering
	\includegraphics[width=0.6\linewidth]{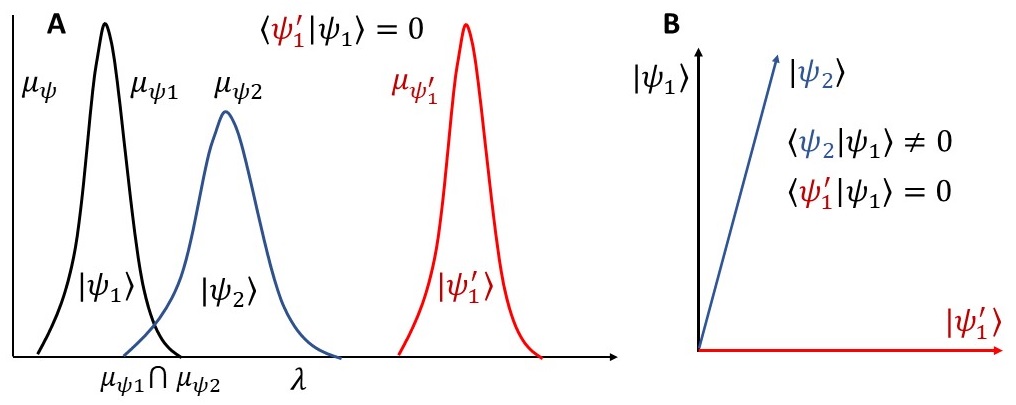}
	\caption{A) Distributions $\mu_{\psi 1}$ and $\mu_{\psi 2}$ with a finite dispersion can overlap with a probability $q$ for some pair of non-orthogonal quantum states $\psi_1$ and $\psi_2$. B) If the physical state $\lambda$ prepared for $\psi_1$ happens to be from the overlap region, a measuring device  associates it with both $\psi_1$ and $\psi_2$ with equal probability (PBR $\psi$-ambiguity). \emph{The results of any measurement should then be identical to those of measuring the state $\psi_2$ with a probability $q/2$}. But this gives a non-zero probability  for the projective measurement relative to a state $\psi'_1$ that is orthogonal to $\psi_1$, showing the inconsistency in the PBR assumption.}
	\label{fig:pbr-inconsist}
\end{figure}

Consider a wavefunction $\psi_1$ and the associated PBR distribution of physical states, with some finite width in the continuous parameter space. A finite non-zero width of the distribution allows a possible finite overlap, quantified by a real positive number `$q$', with some other distribution of physical states corresponding to \emph{another} wavefunction $\psi_2$ (Fig. \ref{fig:pbr-inconsist}A). Since an ideal  device for the preparation of the state $\psi_1$ will be uncertain about the state prepared when the associated physical state $\lambda$ is from the overlap region of the distributions, the prepared physical state corresponds to $\psi_2$ as well, with a probability $q/2$. 
Now consider the inner product $c_o=\langle \psi_1'|\psi_2\rangle\ne 0$. where the state $|\psi_1'\rangle$ orthogonal to $|\psi_1\rangle$; we have $\langle \psi_1'|\psi_1\rangle=0$ (Fig. \ref{fig:pbr-inconsist}B). The  inner product $c_o$ is nonzero because $\psi_2$ is distinct from $\psi_1$ in the linear vector space, and $\langle\psi_1|\psi_2\rangle\ne0$. Therefore, if the starting assumption in the PBR argument is consistent, then a state prepared as $\psi_1$ with an ideal macroscopic device  will give a nonzero number of results when projected to its orthogonal state $|\psi_1'\rangle$, with a probability $q|c_o|^2/2$. This violates the fundamental strict constraint, $\langle \psi_1'|\psi_1\rangle=0$.  Therefore, the PBR assumption of associating a distribution of physical states with a $\psi$-function is inconsistent. 

\subsection{Identical Particles in Quantum Mechanics}
Two-particle wavefunctions of identical particles are of two kinds: symmetric functions representing Bosons and anti-symmetric ones representing Fermions. The `identity' meant here is not merely in the nature of the particle, but in the detailed multi-variable quantum state.

If $|\psi_1\rangle (x_1)$ and $|\psi_2\rangle (x_2)$ are the individual wavefunctions of two identical particles, differing only in the relevant spatial coordinates, $|\Psi_d\rangle=|\psi_1\rangle (x_1) |\psi_2\rangle (x_2)$ is a two-particle wavefunction and $|\Psi_e\rangle=|\psi_1\rangle (x_2) |\psi_2\rangle (x_1)$ is a wavefunction with the coordinates exchanged. Since the quantum state (wavefunction) $\exp(i\phi)|\Psi_e\rangle$ is the same as $|\Psi_e\rangle$, the PBR distributions are the same for both $|\Psi_e\rangle$ and $\exp(i\pi)|\Psi_e\rangle$.

Now consider the symmetric wavefunction 
\begin{equation}\label{key}
	|\Psi_s\rangle=\frac{1}{\sqrt{2}}(|\Psi_d\rangle+|\Psi_e\rangle)=\frac{1}{\sqrt{2}}(|\psi_1\rangle (x_1) |\psi_2\rangle (x_2)+|\psi_1\rangle (x_2) |\psi_2\rangle (x_1))
\end{equation}
The PBR distributions corresponding  to $|\Psi_d\rangle$ and $|\Psi_e\rangle$ are to be combined suitably to make them compatible with $|\Psi_s\rangle$. 

The anti-symmetric wavefunction
\begin{equation}\label{key}
	|\Psi_a\rangle=\frac{1}{\sqrt{2}}(|\Psi_d\rangle-|\Psi_e\rangle)=\frac{1}{\sqrt{2}}(|\psi_1\rangle (x_1) |\psi_2\rangle (x_2)-|\psi_1\rangle (x_2) |\psi_2\rangle (x_1))
\end{equation}  
also needs to be synthesised by combining the same PBR distributions, because the distributions associated with $|\Psi_e\rangle$ and $^-|\Psi_e\rangle$ are identical. But, $|\Psi_a\rangle$ approaches zero for $x_1\approx x_2$. Then, there are no  physical states corresponding to $|\Psi_a\rangle$ when $x_1=x_2$.

\section{Discussion}
\subsection{The Physical States in the Quantum Theory}
It is very important to distinguish between three different elements in the description of dynamics in the quantum theory: the physical system, its dynamical physical state, and \emph{the representation of the physical state in a theory}. A physical system, like a particle with a mass, charge etc. is understood to be `ontologically real' because of the necessity to comply with empirical experience and various conservation constraints. A dynamical physical state  defines the various observables of the system.  In the quantum theory, the representation of the notional physical state is by a $\psi$-function \emph{that specifies (through Born's rule) only the probabilities for the values of the observables as a function of time, and not the values themselves as a function of time}. Statistical results on all observable quantities are obtained from the same  $\psi$-function. Two $\psi$-functions $\psi_1$ and $\psi_2$ represent different physical states only if they differ by more than a pure phase factor $\exp(i\phi)$. Since a quantum state $|s\rangle$ is an element of a linear vector space, it is identical to the representation $\sum_i c_i|r\rangle_i$ 
where $|r\rangle_i$ is any complete basis set corresponding to the eigenstates of some physical observable.  In other words, \emph{both $|s\rangle$ and $\sum_i c_i|r\rangle_i$ represent exactly the same physical state}. The intrinsic statistical dispersion and the uncertainty in quantum mechanics are direct consequences of the equivalence of such multiple representations.  

Notwithstanding any theoretical argument, \emph{we already know empirically that a definite ontological element is related to the wavefunctions of quantum mechanics}. This is because \emph{we are able to alter both the magnitude and the phase of a wavefunction by a purely local interaction with a physical field, a mirror, a waveplate etc.}, all arranged in real space, even when we do not make any measurement. These are local unitary changes of a wavefunction, done in real space and time, with predictable and verifiable outcomes. It is obvious that these local changes, done on individual components of a wavefunction in a situation of multiple possibilities of quantum dynamics, are not done on the physical system (particle), because the physical system itself does not split like a wavefunction. The final observable result depends on the interference of all such accumulated changes of some entity (though directly not the wavefunction itself) that is plurally present, in the multiple paths in real space. Without some entity really propagating in the two paths, this and similar results of single particle interference cannot be explained. \emph{Either one has to propose that it is the wavefunction itself that is ontic, with all its associated problems of state collapse, measurement problem, nonlocality etc., or that there is another more basic dynamical entity with a wave-nature (to be identified in a more fundamental theory) that finally gives rise to an epistemic wavefunction, when ensemble averaged}. That is, an ontic theory of quantum dynamics in which the role of the wavefunction is epistemic, pertaining to a statistical ensemble, is theoretically possible. This view, advocated emphatically by Einstein~\cite{Einstein-Schilpp}, is developed and discussed in ref.~\cite{Unni-UAM}. 

The Bohmian quantum mechanics contains multiple ontic elements that are derived from the abstract wavefunction, like a pilot wave, a nonlocal quantum potential in space, and the spatial trajectories of particles. Though none of these is directly observable, a motivating factor behind the theory was an ontological narrative~\cite{Paavo}. In contrast, the spontaneous collapse models  need in general an ontic existence of (at least) the spatial wavefunctions in the 3-dimensional real space~\cite{Bassi}. However, models that invoke an interaction like gravity between such spatial components of a wavefunction~\cite{Penrose2019,Diosi2021} are  empirically inconsistent, and they are easily refuted by experimental evidence~\cite{UG-gcollapse}.
\subsection{An Analysis of the PBR Argument}
As shown already, the central postulate in the PBR paper, that a distribution of physical states $\mu_\psi(\lambda)$ is associated with a quantum state $\psi$, is inconsistent and invalid when contrasted against the linear structure of quantum mechanics coupled with its quadratic connection to observable results through Born's rule.  Now I review the PBR argument to point out some additional issues arising in the context of  the preparation of quantum states and their projective measurements.

In the PBR scenario, two quantum states $|\psi_1\rangle$ and $|\psi_2\rangle$ are associated with hypothetical distributions of physical states  $\mu_{\psi1}(\lambda)$ and $\mu_{\psi2}(\lambda)$. When a device P for the preparation of a quantum state $|\psi_1\rangle$ is activated, its output is a member $\lambda_1$ from the distribution of the physical states mapped to $|\psi_1\rangle$ (see Fig. 2 in the PBR paper where this is explicit). In other words, what propagates between a device P for state preparation and a device M for measurements is a physical state $\lambda$ (Fig.~\ref{fig:pbr-states}A). This is already  problematic, because if there are multiple paths (possibilities) from P to M, there should be interference effects that decide the exact state reaching M: in fact, the relative phases can make the state reaching M correspond  to the quantum state that is orthogonal to $|\psi\rangle$, in the quantum theory. There is no consistent way of managing this with the real physical states $\lambda$.

The question addressed by PBR was whether two distributions associated with two non-orthogonal $\psi$-functions could have any overlap such that a fraction $q$ of the trials of the state preparation of either quantum state was compatible with both $|\psi_1\rangle$ and $|\psi_2\rangle$. \emph{Thus $q$ represents the ambiguity of $\psi$-functions}, in the PBR model. Quoting from PBR, ``a measuring instrument is uncertain about which state was prepared, and is being projected'', when the physical state $\lambda$ is from the region of overlap. A $\psi$-ontic situation is defined as the one where there is no overlap ($q=0$) for any pair of states with $\langle \psi_1|\psi_2\rangle \ne 0$. (It is obvious that there cannot be an overlap between the distributions when $\langle \psi_1|\psi_2\rangle = 0$.)

The PBR model considers two independent preparation devices A and B, each of which can prepare a physical system in either the quantum state $|0\rangle$ or in the state $|+\rangle$, where $|+\rangle=(|0\rangle+|1\rangle)/\sqrt{2}$. (PBR deal with the general case separately, but this specific case illustrates the argument.) The joint quantum state prepared is one of the four product states $|00\rangle\equiv |0\rangle \otimes |0\rangle$, $|0+\rangle\equiv |0\rangle \otimes |+\rangle$, $|+0\rangle\equiv |+\rangle \otimes |0\rangle$, and $|++\rangle\equiv |+\rangle \otimes |+\rangle$.  Then the states are `brought together' and  measured using projections to specific target quantum states, employing suitable macroscopic devices. PBR assumes explicitly that ``the outcome of a measurement can only depend on the (hypothetical) physical states of the two systems at the time of measurement.'' The target states chosen are four entangled two-particle states, forming an orthogonal basis:
\begin{align}
|\phi_1\rangle=&\frac{1}{\sqrt{2}}(|0\rangle\otimes |1\rangle+|1\rangle\otimes |0\rangle),\hspace{3mm} \nonumber
|\phi_2\rangle=\frac{1}{\sqrt{2}}(|0\rangle\otimes |-\rangle+|1\rangle\otimes |+\rangle) \nonumber\\
|\phi_3\rangle=&\frac{1}{\sqrt{2}}(|+\rangle\otimes |0\rangle+|-\rangle\otimes |0\rangle),\hspace{3mm} 
|\phi_4\rangle=\frac{1}{\sqrt{2}}(|+\rangle\otimes |-\rangle+|-\rangle\otimes |+\rangle) 
\end{align}
PBR notes that the product state $|00\rangle$ is orthogonal to the state $|\phi_1\rangle$, $|0+\rangle$ is orthogonal to $|\phi_2$, $|+0\rangle $ is orthogonal to $|\phi_3$, and 
$|++\rangle$ is orthogonal to $|\phi_4$. Therefore, if the state produced corresponds to $|00\rangle $, say, then the result of the projection to $|\phi_1\rangle$ should be zero. However, if a physical state prepared is from the region of overlap of the distributions, then the pair ($\lambda_1,\lambda_2$) is compatible with all the four quantum states $|00\rangle$, $|0+\rangle$, $|+0\rangle$, and $|++\rangle$. From this PBR state the result: ``This leads immediately to the desired contradiction. At least $q^2$ of the time, the measuring instrument is uncertain  which of the four possible preparation method was used, and on these occasions it runs the risk of giving an outcome that quantum theory predicts should occur with probability $0$.'' Therefore, consistency with the  quantum theory demands no overlap between the distributions.

The same argument of `no overlap' can be repeated for any pair of non-orthogonal states $|\psi_1\rangle$ and $|\psi_2\rangle$. This means that a quantum state can be uniquely inferred from a given physical state $\lambda$. PBR define this as the ontic quantum state. Since $\psi$-epistemic states are defined as those with their associated distributions overlapping, PBR conclude that  $\psi$-epistemic models  contradict the predictions of quantum theory. 

I reiterate that in spite of the statement implying generality, the PBR analysis is not applicable to any $\psi$-epistemic models that treats $\psi$-functions as an element for calculations and predictions, \emph{with no further assumption or speculation about underlying physical states} that are totally irrelevant for quantitative quantum mechanical calculations~\cite{JvN-book,Dirac}. 

Though PBR discussed a projection of the prepared product states to an orthogonal basis set of entangled states, \emph{those product states are actually orthogonal to each component of the entangled state, separately}. For example, the state $|0+\rangle$ is orthogonal to the state $|0\rangle\otimes |-\rangle$ as well to the state $|1\rangle\otimes |+\rangle$ in $|\phi_2\rangle$, \emph{separately}. 

It is important to understand that the measuring device itself is classical and does not represent an entangled quantum state. \emph{If the input two-system quantum state is entangled}, the measuring arrangement can distinguish  the four orthogonal states of an entangled basis. This means that a projective measurement relative to an entangled basis requires entangling the two quantum systems prior to the measurement.  This reveals a conceptual problem in the PBR scenario. Once the states are entangled, neither system corresponds to any $\psi$-function individually; there is only a joint non-separable $\psi$-state for the two particles together. The original mapping between the $\psi$-functions and the distributions of physical states is broken. \emph{The entangling operation itself will be uncertain about which states are involved in the interaction, if the physical state $\lambda$ is from the overlap region of the PBR distributions}. In effect, further assumptions are required about the association of entangles states and distributions of physical states, contrary to the statement by PBR that their assumption needs to hold only for isolated (separable) systems, and not for entangled systems.

\subsection{Additional Remarks}

In the section ``Discussion'' in the PBR paper, a similarity between the PBR result and Bell's theorem is mentioned. They describe Bell's theorem as `no local theory can reproduce the predictions of quantum theory', instead of correctly specifying it as `predictions for quantum correlations cannot be reproduced by a local hidden variable theory (LHVT) of real physical states with definite values for observables'. PBR state that their result is the analogous statement that no $\psi$-epistemic model can reproduce the predictions of quantum theory.  The invalidity of this sweeping assertion is obvious from the fact that the PBR analysis is not applicable to any $\psi$-epistemic model that does not assume distributions of underlying  physical states and such hypothetical connections.  What is striking is the unphysical nature of the central assumptions in both LHVT and the PBR scenarios. For the hidden variable theories that figure in Bell's theorem, the unphysical factor happens to be the gross incompatibility with the fundamental conservation laws~\cite{Unni-JvN}.

PBR extend their discussion to Einstein's restatement of the EPR argument of the lack of one-to-one mapping between a physical state and its $\psi$-representation. They correctly state that Einstein was concerned with the possibility that \emph{there were two or more distinct quantum states for the same real physical state}. However,  the crucial step used to reach this conclusion and Einstein's proof of incompleteness is the fact that the $\psi$-function corresponding to a physical state can be affected and changed by a distant measurement, whereas it is explicitly assumed that the physical state itself cannot be nonlocally affected by such a spatially distant measurement (Einstein locality). This is what ``contradicts the hypothesis of a one-to-one or complete description of the real states'' with the $\psi$-functions in the theory~\cite{Howard}.  What is significant here is the  revealing remark by PBR that the commonly understood notion of `incompleteness' associated with Einstein (erroneously) is a mapping of one $\psi$-function with more than one real physical states.  Though the topic of physical states underlying entangled $\psi$-states is not considered in the PBR paper, leaving it out amounts to a  neglect of a crucial point, especially because PBR use  projections of two-particle states on an entangled measurement basis in their analysis.

In the PBR scenario, there is a mapping between each separable $\psi$-state and a distribution of  physical states. After the two systems interact and become entangled, the joint $\psi$-function should map to a new distribution that reflects the correlations. However, a measurement on one of the systems in any basis that is randomly chosen, at the location A, results in a definite $\psi_A$ that maps to a local distribution. \emph{The same measurement also results in a definite $\psi_B$-function at B instantaneously and nonlocally, which should map to a local distribution $\mu_{\psi_B}(\lambda_B)$}. It is then amply clear that the PBR scenario is not consistent with  a spatially local mapping between $\psi$-functions and distributions of physical states.

\section{Conclusion}
I showed that a widely discussed claim by M. F. Pusey, J. Barrett, and T. Rudolf, of ruling out $\psi$-epistemic models and establishing an ontological status of $\psi$-functions, is flawed. The central assumption in their model, of a one-to-many mapping between a wavefunction and a distribution of underlying physical states, contradicts the linear vector character of the quantum states coupled with Born's rule. I presented several demonstrations of this inconsistency in the context of   both single-particle quantum states and multi-particle states.  The unambiguous conclusion is that the ontological status of the wavefunction cannot be determined by a standard process involving the preparation and projective detection (analysis) of quantum states because none of relevant calculations or statistical results in the standard quantum theory depends on the unknown physical nature of the $\psi$-function. What is ruled out in the PBR analysis is only the specific unphysical model postulated by PBR, and not a possible epistemic nature of the $\psi$-functions of the quantum theory.

\section*{Acknowledgements}
I thank Prof. G. Raghavan for discussions that alerted me about some subtleties, which helped me to sharpen my main argument. I thank Martine Armand for help in the coherent restructuring of the text.

\end{document}